\documentclass{plateau}
\usepackage{listings}
\usepackage{graphicx}
\usepackage{microtype}
\usepackage{xcolor}
\usepackage{circledsteps}

\usepackage{enumitem}
\newlist{enumerate*}{enumerate*}{1}
\setlist[enumerate*,1]{
  label=(\roman*),
  afterlabel=\nobreakspace,
}
\setlist[itemize,1]{
  topsep=4pt,
  leftmargin=2\parindent,
}
\setlist[enumerate,1]{
  label=\arabic*.,
  itemsep=4pt,
  topsep=4pt,
  leftmargin=2\parindent,
}
\setlist[description,1]{
  itemsep=4pt,
  leftmargin=1em,
  topsep=6pt
}

\def\Snospace~{\S{}}

\newcommand\code[1]{\mbox{\texttt{#1}}}

\definecolor{epfl-groseille}{HTML}{b51f1f}
\definecolor{epfl-canard}{HTML}{007480}
\newcommand\tracein{\textcolor{epfl-groseille}{$\mathbf{\Rightarrow}$}}
\newcommand\traceout{\textcolor{epfl-canard}{$\mathbf{\Leftarrow}$}}
\newcommand\tracebar{\textcolor{gray}{|}}

\setcopyright{cc}
\copyrightyear{2026}

\plateauConference[PLATEAU 2026]{16}{March 9--10, 2026}{Pittsburgh, PA, USA}
\plateauDOI{}

\begin{document} 

\title{Tracers for debugging and program exploration}

\author{Shardul Chiplunkar}
\orcid{0000-0002-0803-2133}
\email{shardul.chiplunkar@epfl.ch}

\author{Clément Pit-Claudel}
\orcid{0000-0002-1900-3901}
\email{clement.pit-claudel@epfl.ch}

\affiliation{%
  \department{School of Computer and Communication Sciences}
  \institution{EPFL}
  \city{Lausanne}
  \country{Switzerland}
}

\begin{abstract}
  Programmers often use an iterative process of hypothesis generation (``perhaps this function is called twice?'') and hypothesis testing (``let's count how many times this breakpoint fires'') to understand the behavior of unfamiliar or malfunctioning software.
  Existing debugging tools are much better suited to testing hypotheses than to generating them.
  Step debuggers, for example, present isolated snapshots of the program's state, leaving it to the programmer to mentally reconstruct the evolution of that state over time.
  We advocate for a different approach:
  building a debugging and program-exploration tool around a \emph{trace}, or complete history, of the program's execution.
  Our key claim is that the user should see every line \emph{as executed} (in time order) rather than \emph{as written} (in syntax order).
  We discuss design choices, preliminary results, and interesting challenges.
\end{abstract}

\keywords{tracing, program visualization, programming tools}

\maketitle

\begin{figure}[h]
  \centering
  \vspace{-4mm}
  \includegraphics[width=\linewidth]{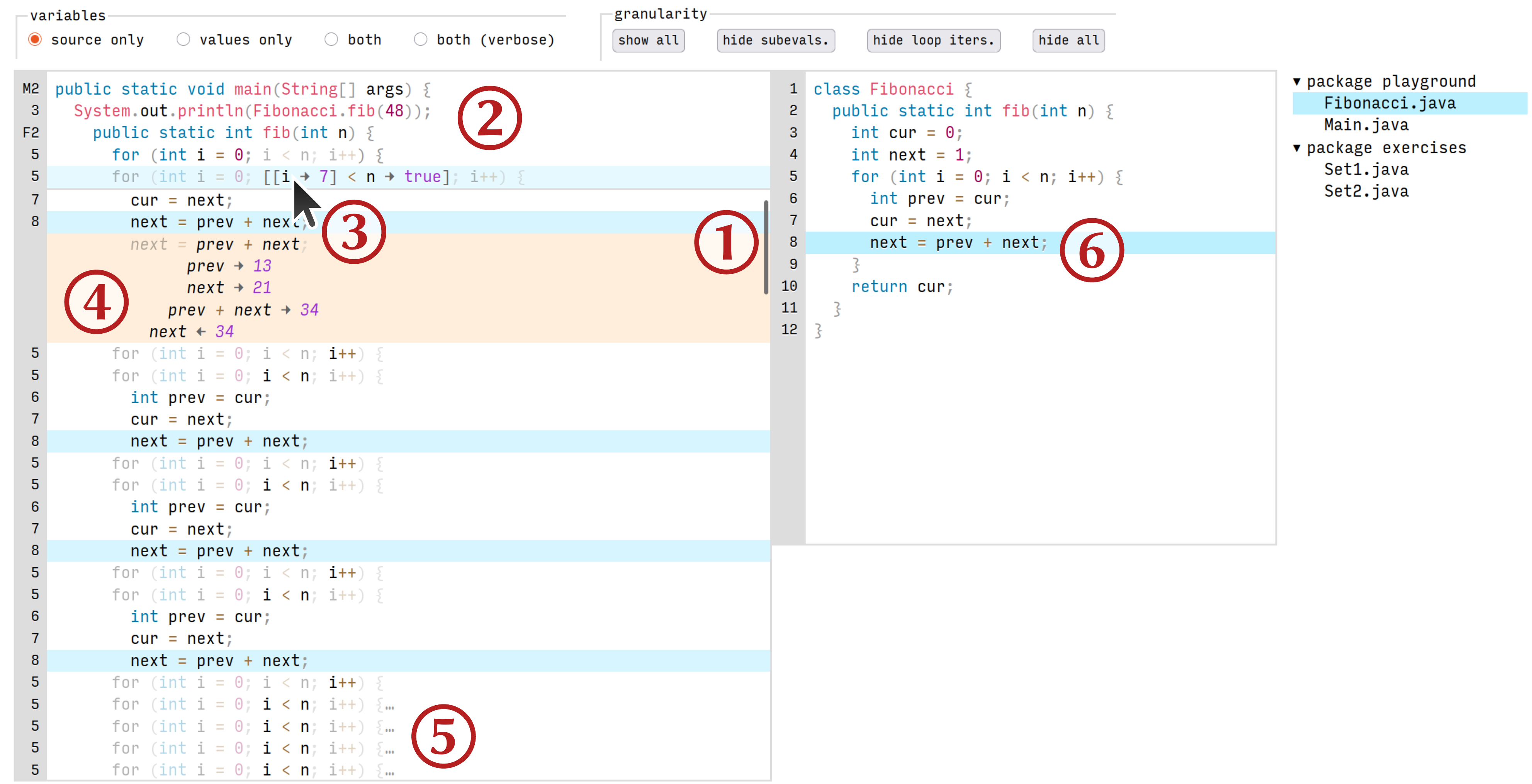}
  \caption{%
    Mockup of our proposed tracer.
    On the left is the primary trace interface, and on the right, the traditional source code.
    \Circled{\textbf{1}}~In lieu of stepping, the user freely navigates the trace by scrolling, but can also select a line in the trace and navigate by keyboard if desired.
    \Circled{\textbf{2}}~As they scroll, a ``stack trace'' of method and loop headers sticks to the top of the buffer.
    \Circled{\textbf{3}}~Hovering over an expression in the trace while pressing \texttt{Ctrl} shows its value and the values of any containing expressions.
    \Circled{\textbf{4}}~Subexpression evaluations can alternatively be shown by expanding a statement, and \Circled{\textbf{5}}~loop and method blocks can be collapsed, letting the user control the display granularity.
    \Circled{\textbf{6}}~Selecting a line in the source code highlights all instances of its execution in the trace, and \emph{vice versa}.
    (Pointer icon from Ubuntu Yaru theme, CC-BY-SA.)
  }
  \label{fig:teaser}
\end{figure}

\section{Introduction}

When a programmer is \emph{debugging} a program, they are applying the scientific method to understand where and why its observed behavior diverges from their expectations:
starting from a working theory of the program’s behavior, they generate a hypothesis, design an experiment that could support or falsify it, conduct the experiment, incorporate the result into their theory, and repeat, until their theory can explain the observed (unexpected) behavior \cite{ZellerWhyProgramsFail2009,AlaboudiLaTozaHypothesizerHypothesisBasedDebugger2023}.
At that point, they have ``found the bug'' and can start trying to fix it.
For example,
\begin{itemize}
\item
  The working theory (based on previous experiments) might be that the observed wrong output is caused by a wrong counter value.
\item
  A hypothesis might be, ``The \code{set\_counter} function is sometimes called with argument \code{value = 0} when it should always be called with \code{value = 1}''.
\item
  A corresponding experiment might be to add the statement \code{print(value)} right at the start of the definition of \code{set\_counter} and rerun the program, or set a breakpoint on \code{set\_counter} in a debugger and inspect \code{value}.
\item
  If the outcome upholds the hypothesis, i.e.~the program's behavior has already diverged from expectations before the observation, then the next hypothesis would probably concern the call sites of \code{set\_counter};
  else it has not diverged yet, and the next hypothesis might concern the usage of \code{value} in the method body and further calls.
\end{itemize}

Beyond debugging, the process above also describes program \emph{exploration}.
A programmer looking to understand an unfamiliar codebase (say, in order to add a feature, or make up for a lack of documentation) may engage in a similar process, as may a student trying to learn a language by studying (not necessarily buggy) example programs.
In each case, the user incrementally builds a theory of the program's behavior based on dynamic observations in addition to static source code.

The two techniques mentioned above---using breakpoints in a debugger, and \emph{\code{printf}-debugging} by adding \code{print} statements (or similar)---are well-suited for \emph{testing} hypotheses.
Simple hypotheses about the values of variables, the order of execution of statements, etc.\ are easy to test with either technique, and more complex ones are possible with some forethought.
For example, to investigate the recursion depth of a function, a programmer might set a breakpoint on the base cases and inspect the call stack, or augment the function with a counter argument incremented on every call and printed out from the base cases.

However, neither breakpoints nor print statements directly help the programmer \emph{develop} hypotheses about the program.
If the program seems to run indefinitely, is it because of an infinite loop, or a blocking call to an external resource, or a very expensive computation, or something else?
The programmer must rely on their intuition to first choose one of these potential causes to investigate, and to decide where in the code to investigate it, before a breakpoint or print statement can be of help in confirming or refuting the hypothesis.

\begin{figure}[t]
  \centering
  \begin{minipage}[t]{0.35\linewidth}
\begin{lstlisting}[%
escapechar=!,
columns=flexible]
# file: logic.py
def do_it():
  args = read_from_file()
  result = !\textcolor{epfl-groseille}{compute(args)}!
  return args, result

def compute(things):
  !\textcolor{epfl-canard}{\textbf{return} \textbf{sum}(things)}!


# file: main.py
def main():
  initialize()
  result, _ = !\textcolor{epfl-groseille}{do\_it()}!
  process(result)

!\textcolor{epfl-groseille}{main()}!
\end{lstlisting}
  \end{minipage}\qquad%
  \begin{minipage}[t]{0.55\linewidth}
\begin{lstlisting}[%
escapechar=!,
columns=flexible]
main():
!\tracebar! initialize() !\textcolor{gray}{[$\,\dotsc$]}!
!\tracebar! result, _ = do_it()
!\tracebar!   do_it():
!\tracebar!   !\tracebar! args = read_from_file()
!\tracebar!   !\tracebar!   read_from_file() !\traceout! [2, 3, 5]
!\tracebar!   !\tracebar!   args !\tracein! [2, 3, 5]
!\tracebar!   !\tracebar! result = compute(args !\traceout! [2, 3, 5])
!\tracebar!   !\tracebar!   compute(things !\tracein! [2, 3, 5]):
!\tracebar!   !\tracebar!   !\tracebar! return sum(things !\traceout! [2, 3, 5])
!\tracebar!   !\tracebar!   !\tracebar!   sum(things !\tracein! [2, 3, 5]) !\traceout! 10
!\tracebar!   !\tracebar!   !\tracebar! !\traceout! 10
!\tracebar!   !\tracebar!   result !\tracein! 10
!\tracebar!   !\tracebar! return args, result
!\tracebar!   !\tracebar! !\traceout! [2, 3, 5], 10
!\tracebar!   result !\tracein! [2, 3, 5]
!\tracebar! process(result !\traceout! [2, 3, 5]) !\textcolor{gray}{[$\,\dotsc$]}!
\end{lstlisting}
  \end{minipage}
  \caption{%
    On the left is a program stopped at \lstinline{return sum(things)} (in blue) inside a stepper.
    The stack trace, consisting of \lstinline{compute}, \lstinline{do_it}, and \lstinline{main}, is highlighted in red;
    but as \lstinline{main} is in a different file, the stepper would not display its source code at the same time as that of the other two functions.
    On the right is the same program in a tracer.
    (It is not stopped anywhere;
    it has run to completion.
    If it had stopped, say, due to an error while computing \lstinline{sum(things)}, the trace would have been truncated at that point.)
    Red left arrows mark values passed into function calls or assigned to variables, while blue right arrows mark returned or referenced values.
  The bodies of \lstinline{initialize} and \lstinline{process} have been collapsed by the user.}
  \label{fig:side-by-side}
\end{figure}

But what can help a programmer develop hypotheses is observing, \emph{over time}, a collection of breakpoints or print statements, i.e.~\emph{stepping} through the program in the debugger or \emph{logging} key points in its execution.
Rather than probing at a specific point in time, a programmer might use these techniques to get an overview of the whole execution, and perhaps try to notice where it diverges from their expectations.
They come at higher cost, though:
step-debugging requires careful planning (conditional breakpoints, and stepping ``over'', ``into'', ``out of'', etc., often with no ``undo'') to navigate to the relevant part of the execution and is tricky to reproduce;
logging requires manually inserting logging statements and investing effort in making their output useful (selecting what and where to log, indenting or otherwise structuring the log, etc.).

Based on these observations, we propose a new kind of tool for program exploration and debugging—a \emph{tracer}—that helps the user develop hypotheses about a program.
A tracer generates, and then lets the user browse and query, a complete trace of a program's execution on a given input, recording every statement in order of execution along with the values of any variables and other expressions it contains.
\autoref{fig:teaser} shows a web-based mockup we have been working on (discussed in~\autoref{sec:current-progress}), while \autoref{fig:side-by-side} shows a more barebones, plain-text example.
Crucially, when the user browses the trace, the \emph{y}-axis represents time:
the statement preceding (following) a given statement is the one that executed just prior (just after), or in other words, code is shown in its \emph{temporal context}.
This is in contrast with a stepper, which shows code in its \emph{lexical context} and uses real-world time to represent time, controlled with step actions.
This also differs from a logger in displaying code instead of custom messages, although both are in temporal order.

In the rest of this paper, we will first describe the limitations of existing stepping and logging tools, and of those approaches in general.
Then, we will sketch how a tracer might address those limitations to enable a new mode of exploring and debugging programs.
Next, we will briefly report on our ongoing work towards building a practical tracer.
Finally, we will conclude with a discussion of related prior work.

\section{Limitations of existing tools and approaches}
\label{sec:limitations}

Existing tools for debugging and program exploration can be classified into \emph{steppers} and \emph{loggers}, explained in the subsections below.
Both categories have their limitations.
Some are inherent to the approach, whereas others could be remedied by an implementation but commonly aren't.

\subsection{Steppers}
\label{sec:limitations:steppers}

In this section, ``steppers'' refers to tools that let the user step through the execution of a program, whether or not for debugging purposes.
In addition to the source code and the current locus of execution, they may display information such as the current stack trace, the values of local variables, or the standard output and error streams.

\begin{description}
\item[Lexical context-switching obscures temporal continuity.]
  By design, a stepper shows the current line of code in its lexical context, limited by screen space to typically just the enclosing function definition.
  Thus, any action that changes the lexical context, such as stepping into or returning from a function call, changes the display entirely and requires the user to context-switch to a different location in the source code---potentially losing track of the temporal continuity of the computation.
  To reconstruct the temporal context of the current line, i.e.~what computations have just happened and what are to happen immediately afterwards, the user must navigate to (or simply remember) previous stack frames.

\item[Mistakes are costly, inhibiting exploration.]
  Many steppers cannot fully undo all actions.
  To restore the previous state of the stepper, the user must restart it and tediously redo their sequence of actions that they must have already kept track of separately.
  This process can sometimes be sped up by more complex actions like setting conditional breakpoints and watchpoints if the stepper supports them and the user knows how to use them, at the potential additional cost of making and debugging a meta-level mistake (e.g.~an incorrect condition).
  As a result, a user is more likely to think carefully before taking an action;
  using the stepper feels more effortful and less exploratory.

\item[Revisiting the past is costly.]
  Even without making mistakes, the common lack of backwards actions makes it difficult to revisit the past of a computation.
  Consider a user who, using a stepper, has just confirmed their hypothesis about an erroneous value at some line of code, and concluded that the program has already diverged from their expectations before that line.
  To test a new hypothesis earlier in the program's execution, they must often simply restart the stepper, suffering similar costs as described before.

\item[Stack traces omit looping information (unlike recursion).]
  A stepper that displays stack traces will show the preceding sequence of recursive calls when stopped inside a recursive function;
  whereas, inside a loop, it won't provide any information about previous iterations of the loop.
  The disparity is even more evident between a tail recursive function and its loop counterpart, where the two computations have the same structure, and may even be identical after optimization.


\end{description}

\subsection{\code{print}, and other loggers}
\label{sec:limitations:loggers}

In this section, ``loggers'' refers broadly to any means of recording information at user-chosen points in the program's execution, from \code{print} statements to full-featured logging libraries.

\begin{description}
\item[Setup cost grows with program size.]
  The setup cost of logging includes the cost of writing logging statements with all relevant information at all possible places of interest in the program, which scales with the size of the program, in addition to the one-time cost of setting up the logger itself.
  This is one of the main reasons a user may choose to switch to a stepper after initial exploration or debugging with simple \code{print} statements:
  it is hard to judge whether investing in more structured output from more places in the program will be worth it, or whether they will find what they're looking for in a few iterations of using the stepper.

\item[Implicit control flow is hard to understand or query.]
  Logging records the sequence of actions a program takes but often does not explicitly record control flow.
  For instance, in \autoref{fig:side-by-side}, suppose \lstinline{do_it} logged a message like ``computing the result with arguments <args>'' just before calling \lstinline{compute}, and \lstinline{compute} logged ``computation took <t> seconds'' just before returning, and the logger annotated all messages with the name of the function that produced them.
  A user reading the log must infer that a function call or return occurred between the two messages from the fact that they are annotated with different function names, and must rely on external information (e.g.~the source code) to disambiguate the two possibilities.
  (The user could, of course, choose to log every call, branch, loop, etc.\ at significant additional setup cost.)
  As a corollary, being implicit makes control flow hard to query.
  Continuing the example above, filtering the log to only show output from \lstinline{compute} when called from \lstinline{do_it} is tricky at best and not unambiguously possible at worst.
\end{description}

\section{The ideal tracer}
\label{sec:ideal}

Recall that our proposed tracer generates, and then lets the user browse and query, a complete trace of a program's execution on a given input, recording every statement in order of execution along with the values of any variables and other expressions it contains.
See \autoref{fig:teaser} and \autoref{fig:side-by-side} for mocked-up examples.
Such a tracer ideally combines the best of steppers and loggers to address the limitations above.
Again, some of these improvements are inherent to tracing, and others are enabled by a particular implementation or user interface affordance.

\begin{description}
\item[Code is displayed in its temporal context with concrete values.]
  By using the \emph{y}-axis to represent the flow of time during a program's execution, tracing spares the user the cognitive effort of keeping track of the temporal context across function calls, loops, and other control flow events.
  Meanwhile, by annotating variables in the trace with their concrete values, tracing retains the benefits of inspecting them in a stepper and makes it easier to follow how data flows through the program.
  Note that the stack trace at any point can be trivially recovered from the trace as the sequence of enclosing function calls.

\item[Mistakes are disconnected from program state, hence easy to revert.]
  By the time a user is interacting with the tracer, the program has already finished running, and its state has been recorded at each point in its execution;
  there is no ``current program state'' that could be irreversibly modified by the user's mistakes.
  The user only manipulates their view of the trace and associated information, and undoing those actions is a simpler user-interface concern.

\item[To visit the past (or future), the user simply scrolls up (or down).]
  Replacing the fixed basic actions of stepping backwards and forwards with the flexible ones of scrolling up and down makes iterated exploratory use easier.

\item[Looping is as salient as recursion.]
  Tracing records every executed statement, whether for looping or recursion.
  A tracer may choose to \emph{display} the two differently---for instance, displaying iterations of a loop at the same indentation level, but increasing it for recursive calls---but still presents the user with equal amounts of information.

\item[Setup cost is independent of program size.]
  Tracing is automatic, and like using a stepper, only requires rerunning (after potentially recompiling) the program, without manually modifying it.

\item[Control flow is explicit, hence easier to understand or query.]
  Tracing records every executed statement, including control flow statements, so the trace contains complete, explicit information about the control flow of the program.
  This makes queries like ``show output from \lstinline{compute} when called from \lstinline{do_it}'' possible (but how easy they are to implement can depend on the specific trace format).
  Plus, a tracer displays program execution as code rather than as messages about the code, making the control flow more directly evident to the user.

\end{description}

\subsection{Common concerns}

There are a number of natural concerns about such a tool, which we address below.

One is that recording, displaying, and querying the trace would present too high a performance overhead to be practical for programs of realistic sizes.
To allay this concern, we first note that we see tracers primarily being useful at human scale:
individual programmers writing scripts for their own use, a small team building a research prototype, students working on programming assignments.
(Multi-threaded or distributed programs are outside our scope.)
In these contexts, ``realistic'' programs rarely exceed a few thousand lines of code, run for more than a few seconds, or use more than a narrow subset of the capabilities of external libraries, if at all.
A tracer would not need to scale further to be useful, echoing the philosophy of \citet{KangGuoOmnicodeNoviceOrientedLive2017}.
Moreover, a back-of-the-envelope calculation suggests that a Python interpreter that executes $10^6$~lines of code per second, running for 2~s, producing 1~kB of trace data per line of code, would produce 2~GB of trace data---while standard stream utilities like \code{grep} can process 1~GB of data per second, i.e.~can perform simple queries on the trace in roughly the same amount of time it took to generate it.
A proper overall evaluation of our tracer would of course include a performance evaluation.

Another related concern is that the trace would overwhelm the user with information.
Pending empirical evaluation, we believe some simple features for the tracer's user interface would go a long way in addressing this concern:
\begin{itemize}
\item
  \emph{Structural queries} about the trace, letting the user search for control flow patterns alongside identifier names and concrete values.
  For instance, searching for the first iteration of a given loop in which a given variable is assigned the value \code{null}; or searching for references to a given variable inside a given recursive function, but not inside its calls to other functions.
  A drag-and-drop interface could let users compose queries out of basic building blocks, internally representing the query in a tree-based selector language like CSS or XPath that advanced users could edit directly.
\item
  Standard browsing utilities like bookmarks (at points of interest in the trace), saved searches, collapsing/expanding conceptual units to control granularity (like function calls or iterations of a loop; see \Circled{\textbf{5}} in \autoref{fig:teaser}), contextual breadcrumbs (see \Circled{\textbf{2}}), and coarse-grained scrolling through an overview or ``minimap'' of the trace.
\item
  Click-to-inspect records and objects, also letting the user track object identity through the trace.
  (Primitive values can always be displayed as inline annotations; see \Circled{\textbf{3}} and \Circled{\textbf{4}} in \autoref{fig:teaser}.)
\end{itemize}
These features can be seen as instances of \citet{VictorLearnableProgramming2012}'s principles for a ``learnable programming'' environment, that lets users ``follow the flow'', ``see the state'', and ``create by reacting'' (in our case, to ``create'' would be to refine hypotheses).

A third concern is that users still need lexical context to understand a program.
This is easily remedied by showing the source code in a secondary pane beside the trace, and letting the user select a line in the source code to see all instances of its execution in the trace (e.g., line~8 is selected in \Circled{\textbf{6}} in \autoref{fig:teaser}), or conversely, select a line in the trace to bring up its lexical context.
Whether having this information helps users more than it distracts them could be an object of evaluation.

Lastly, the description of a program in terms of executed statements and control flow decisions leans imperative in style, and may not work as well for tracing functional code.
Functional programs typically have more nested subexpressions, more anonymous intermediate values, fewer named variables, and fewer top-level statements than imperative programs.
Yet, it is functional languages that more commonly already have tracing utilities, such as \code{\#trace} in OCaml~\cite{INRIAToplevelSystemREPL2025} and \code{trace} in Common Lisp~\cite{LispWorksLtd.MacroTRACEUNTRACE2025}, often limited to function calls and return values instead of tracing all subexpressions.
This may be sufficient to make tracing useful in practice for those languages, while imperative-style languages might need more detailed traces.

\subsection{Current progress}
\label{sec:current-progress}

Several student projects \cite{AebiImmediateTracing2023,KappelerPrintWizardNextlevelTracebased2026,JolidonKappelerStudentreadyImplementationTracebased2025,SerandourImmediateTracingSmoother2024} that the authors supervised have made initial progress on recording traces of Java programs,
along with proof-of-concept user interfaces for browsing those traces.
The authors' work on a full-featured user interface is in early stages and is shown as a mockup in \autoref{fig:teaser}.

\section{Related work}

Tracing and related ideas have a long history.
We only summarize the most relevant prior work below, drawing from the terminology and works surveyed in \citet{EngblomReviewReverseDebugging2012}, \citet{HeinsenEganAdvancedDebuggingProgram2015}, and \citet{LamperthRecordReplayDebugging2024}, which are good resources for further reading.

\subsection{Enhanced steppers and loggers}

Noting many of the same problems with steppers and loggers as we outlined in \autoref{sec:limitations}, some prior work has focused on enhancing those tools while maintaining their basic principles of operation.

\emph{Record and replay} techniques focus on determinizing program behavior by recording or controlling interactions with sources of nondeterminism, like thread scheduling or timers, during one run of the program to allow exactly identical reruns.
These recordings support standard step-debugging and sometimes also backwards stepping or other debugger actions.
Perhaps the best known example is the rr~project and its commercial frontend Pernosco, focusing on languages like C, C++, Ada, Rust, and V8~JS \cite{OCallahanEtAlEngineeringRecordReplay2017,PernoscoPernosco2026};
record-and-replay debuggers for JVM-based languages include \citet{HuangEtAlLEAPLightweightDeterministic2010}, \citet{LamperthRecordReplayDebugging2024}, and \citet{SchwartzEtAlJmvxFastMultithreaded2024}.
The aims of record-and-replay differ from tracing in two important ways.
One, replay is designed to occur primarily (often only) inside a stepper with a live instance of the program, even though the recording may contain enough information to reconstitute a trace.
Two, the nondeterminism that the technique shines in debugging most often occurs in large-scale programs or whole systems, so the projects above make tradeoffs in favor of performance over other goals like explorability.


\emph{Omniscient} debugging consists of recording many more kinds of information during the execution of a program to extend the capabilities of a standard stepper, such as investigating value provenance.
The term was coined by \citet{LewisDebuggingBackwardsTime2003}, but the idea goes back to the ExDAMS project \cite{BalzerExDAMSExtendableDebugging1969}.
The information collected is often called a ``trace'', but is not displayed as such, rather serving as a database for a step-based interface, which is again a key difference from our notion of tracing.
Notable omniscient debuggers include Zstep for Lisp, whose design started from premises strikingly similar to ours \cite{LiebermanStepsBetterDebugging1984};
the pedagogical Execution Trace Viewer (ETV), which featured a uniform (internal) trace format for C, Java, Perl, and UtiLisp \cite{TeradaETVProgramTrace2005};
the Trace-Oriented Debugger (TOD) for Java \cite{PothierTanterBackFutureOmniscient2009};
commercial products like IntelliTrace \cite{MicrosoftCorp.IntelliTrace2025}, Undo \cite{UndoLtd.UndoJava2026}, and Replay \cite{RecordReplayInc.Replay2026};
and the popular online educational tool PythonTutor, eventually extended to Java, JavaScript, C, and C++, and still in active use today \cite{GuoOnlinePythonTutor2013,GuoTenMillionUsers2021}.
Meanwhile, \citet{WangLaTozaHowOmniscientDebuggers2025} question whether omniscient debuggers really have productivity benefits, and identify some of the same obstacles as we do in~\autoref{sec:limitations:steppers} through a user study.


Enhanced logging has been less extensively studied.
\citet{JiangEtAlLogItSupportingProgramming2023} build Log-It for JavaScript, a logger that adds more structure and information to log output without being too costly for developers to use.
Their design goals and functionality are closely aligned with our proposed tracer.
A more limited approach, which is more common in the literature, focuses on automatically inserting or modifying simple log statements, such as in \cite{KoEtAlAutoPrintJudgingEffectiveness2025};
see, e.g., \citet{ZhongEtAlLogUpdaterAutomatedDetection2025} for further references.
Lastly, \citet{BianchiEtAlInlineVisualizationManipulation2024} develop an expression language for filtering and manipulating log output containing runtime values for inline display in the source code.

\subsection{Tracers}

Unlike the tools in the previous section, \emph{tracing} centers the trace:
the primary mode of interaction with a tracer is directly browsing and querying a complete trace of a program’s execution, which includes other recorded information like the values of variables.
We found five prior works that approach tracing in this sense for imperative languages:
CMeRun for C++ \cite{EtheredgeCMeRunProgramLogic2004};
Backstop \cite{MurphyEtAlBackstopToolDebugging2008}, Traceglasses \cite{SakuraiEtAlTraceglassesTracebasedDebugger2010}, and Trace Debugger \cite{ArtiukhovEtAlTraceDebuggerInteractive2026} for Java;
and snoop \cite{HallSnoop2024} for Python (succeeding PySnooper \cite{RachumEtAlPySnooperNeverUse2019}).
CMeRun and Backstop programmatically insert print statements after every line of the original program that print the line of code that was just executed along with concrete variable values to standard output.
Compiling and running this instrumented program produces a plain-text trace that both tools display directly.
Traceglasses presents the trace in a UI that allows for easier browsing and searching, but the trace does not always strictly correspond in syntax or structure to the source code, as it is the result of instrumenting the JVM bytecode instead.
The Trace Debugger (also instrumenting JVM bytecode) perhaps comes closest to our goals, but still does not fundamentally depart from the stepper paradigm: the trace only seems to show lines of code corresponding to control flow events, instead of every line executed, and navigating program execution via the trace serves mainly to augment the standard step navigation built around a source-code view.
Lastly, snoop hooks into Python's built-in \texttt{sys.settrace} method to produce a plain-text trace with concrete values, and offers a customization API and a GUI to inspect values from different function calls and loop iterations while browsing the source code \cite{HallBirdseye2025}.


For functional programming languages, in which a program's execution consists mostly of applying and evaluating functions, the most natural form of a trace is a call tree, which serves most of the same purposes.
Call-based tracing has historically been built into some languages like Common Lisp \cite{LispWorksLtd.MacroTRACEUNTRACE2025} and OCaml \cite{INRIAToplevelSystemREPL2025}, and has been developed separately for others, including Clojure \cite{MonettaFlowStorm2025} and a subset of Haskell \cite{VasconcelosHaskelite2025,PosmaKrouseLessons2014}.
\citet{KleynGingrichGraphTraceUnderstandingObjectoriented1988}, one of the oldest discussions we found of directly visualizing execution traces, generalize call trees to call graphs in the context of object-oriented programming in Lisp, allowing users to explore individual objects, classes, methods, and their relationships.
\citet{HoferEtAlDesignImplementationBackwardintime2006} describe a similar tool for Smalltalk.
For imperative languages, call trees capture more limited information about the program's behavior but may still be valuable, and are sometimes displayed alongside stepper interfaces, as in TOD for Java \cite{PothierTanterBackFutureOmniscient2009}, Theseus for JavaScript \cite{LieberTheseus2015}, and the pedagogical JavaWiz \cite{WeningerEtAlJavaWizTracebasedGraphical2025} and SRec for Java \cite{Velazquez-IturbideEtAlSRecAnimationSystem2008}.
Lastly, tracing takes a rather different form for logic programming languages, such as in \citet{EisenstadtBrayshawTransparentPROLOGMachine1988} or \citet{SWIPrologDebuggingTracingPrograms2026}.

\subsection{Trace-based analyses}

Beyond human-facing tracers, execution traces can power analyses like finding anomalous program executions and their causes. We briefly mention key research below.

Traces can be used to assist programmers in generating hypotheses.
One approach is with ``interrogative debugging'', where a programmer can ask ``why did'' or ``why didn't'' questions about the data or control flow of the program to refine their hypotheses \cite{KoMyersDesigningWhylineDebugging2004}.
The Whyline, an interrogative debugger for Java, analyzes execution traces to answer these questions \cite{KoMyersDebuggingReinventedAsking2008}.
Another approach is ``hypothesis-based debugging'', taken by Hypothesizer \cite{AlaboudiLaTozaHypothesizerHypothesisBasedDebugger2023}, where a candidate hypothesis is automatically drawn from a pre-specified database of possible hypotheses by matching their conditions against the observed trace.

\emph{Trace diffing}---comparing two execution traces of the same program, typically a successful and an unsuccessful one---can help locate bugs.
Explored in detail by Zeller in his work on Delta Debugging (e.g, \cite{CleveZellerLocatingCausesProgram2005}) and more generally in the literature on \emph{path profiling} \cite{RepsEtAlUseProgramProfiling1997}, trace diffing continues to be relevant in contexts ranging from programming coursework feedback \cite{SuzukiEtAlTraceDiffDebuggingUnexpected2017} to trigger-action programming for smart devices
\cite{ZhaoEtAlUnderstandingTriggerActionPrograms2021}.

Trace diffing and path profiling are a special case of \emph{program spectra} analysis \cite{HarroldEtAlEmpiricalInvestigationRelationship2000}.
A spectrum is a ``signature'' of a program's execution that can be as simple as the set of conditional branches that were ever executed, or as complete as a full trace.
A large body of work has investigated both the analysis of spectra (such as \cite{JonesHarroldEmpiricalEvaluationTarantula2005,WongEtAlDStarMethodEffective2014,AlimadadiEtAlInferringHierarchicalMotifs2018}) and their presentation to the user (such as \cite{CornelissenEtAlControlledExperimentProgram2011,MatsumuraEtAlRepeatedlyexecutedmethodViewerEfficient2014,LieberEtAlAddressingMisconceptionsCode2014}).


\subsection{Querying traces}

In~\autoref{sec:ideal}, we mentioned that the ideal experience of using a tracer should include the ability to query the structure (control flow) as well as the content (values) of the trace.
Past work has considered the design of such query languages.
\citet{MartinEtAlFindingApplicationErrors2005} describe a query language for Java programs focused on objects and state, and implement an overapproximative static checker to reduce the amount of dynamic checking required;
\citet{GoldsmithEtAlRelationalQueriesProgram2005} present an SQL-inspired query language focused on temporal relationships between invocations and allocations;
\citet{ConsensEtAlVisualizingQueryingDistributed1994} develop a visual, graph-based query tool for ``distributed event traces'', a different notion of trace;
and similarly, \citet{LeDouxParkerSavingTracesAda1985a} also focus on distributed communication, but their use of Prolog as the query language enables more general trace queries, such as ``qualify[ing] a query with temporal constraints to simulate a breakpoint retroactively''.
All of these query languages operate over a relatively flat collection of (possibly timestamped) events, whereas our vision of a trace is fundamentally structured as a tree corresponding to the program's control flow.

\subsection{Live, literate, and learnable programming}

Our proposed tracer is diametrically opposed to live programming, dealing exclusively with ``dead'' programs that have already finished executing, and also to literate programming, instead only showing lines of the source code with minimal annotations.
Nonetheless, those techniques embody aspects of \emph{learnable programming}---bringing the dynamic behavior of a program closer to mental models that are more intuitive for the programmer, rather than making the programmer compensate for the limitations of the machine or environment \cite{VictorLearnableProgramming2012}---that certainly bear upon our work.

Specifically, a tracer surfaces otherwise hidden runtime state to the user.
Starting from similar observations as ours concerning the limitations of steppers and loggers, Omnicode, a Python programming environment, ``push[es] live programming [\ldots] to the extreme by displaying the entire history of all run-time values for all program variables all the time'' in the form of a scatterplot matrix \cite{KangGuoOmnicodeNoviceOrientedLive2017}.
The authors show that this approach is practical and useful for novices writing self-contained algorithmic programs.
Alternatively, \emph{projection boxes} allow for inline annotations of runtime values or other information in a program's source code \cite{LernerProjectionBoxesOnthefly2020}.
A different domain where runtime state is essential to understanding the program is with interactive theorem provers:
Proviola~\cite{TankinkEtAlProviolaToolProof2010} records and displays the output of the prover, showing the proof state at every step of the proof, alongside the input proof script.

Literate programming extends the idea above to interspersing prose explanations with code, often containing example evaluations or visualizations of dynamic behavior.
Alectryon~\cite{Pit-ClaudelUntanglingMechanizedProofs2020} applies this idea to proof scripts for interactive theorem provers;
\citet{SotoudehLiterateTracing2025} to traces of program execution, but in the form of interactive steppers and custom visualizations;
and Jupyter~\cite{KluyverEtAlJupyterNotebooksPublishing2016} to computational scientific documents or ``notebooks''.
``Literate traces'' could likewise be a promising means for communicating dynamic program behavior.




\begin{acks}
  We would like to thank our PLATEAU mentors, Hila Peleg and Joe Gibbs Politz, as well as all the attendees of PLATEAU~2026, for their helpful feedback, abundant pointers to the literature, and enthusiasm about this work.
\end{acks}

\bibliography{refs}

@techreport{AebiImmediateTracing2023,
  type = {Semester Project},
  title = {Immediate Tracing},
  author = {Aebi, Valentin},
  year = 2023,
  month = jul,
  address = {Lausanne, Switzerland},
  institution = {EPFL},
  url = {https://infoscience.epfl.ch/handle/20.500.14299/198892},
  urldate = {2026-04-07},
  langid = {english}
}

@inproceedings{AlaboudiLaTozaHypothesizerHypothesisbasedDebugger2023,
  title = {Hypothesizer: A Hypothesis-Based Debugger to Find and Test Debugging Hypotheses},
  shorttitle = {Hypothesizer},
  booktitle = {Proceedings of the 36th {{Annual ACM Symposium}} on {{User Interface Software}} and {{Technology}}},
  author = {Alaboudi, Abdulaziz and LaToza, Thomas D.},
  year = 2023,
  month = oct,
  series = {{{UIST}} 2023},
  pages = {1--14},
  publisher = {ACM},
  address = {New York, NY, USA},
  doi = {10.1145/3586183.3606781},
  urldate = {2026-02-10},
  isbn = {979-8-4007-0132-0},
  langid = {english}
}

@inproceedings{AlimadadiEtAlInferringHierarchicalMotifs2018,
  title = {Inferring Hierarchical Motifs from Execution Traces},
  booktitle = {Proceedings of the 40th {{International Conference}} on {{Software Engineering}}},
  author = {Alimadadi, Saba and Mesbah, Ali and Pattabiraman, Karthik},
  year = 2018,
  month = may,
  series = {{{ICSE}} 2018},
  pages = {776--787},
  publisher = {Association for Computing Machinery},
  address = {New York, NY, USA},
  doi = {10.1145/3180155.3180216},
  urldate = {2026-02-10},
  isbn = {978-1-4503-5638-1}
}

@article{ArtiukhovEtAlTraceDebuggerInteractive2026,
  title = {Trace Debugger: Interactive Execution Trace Debugging for {{Java}} and {{Kotlin}}},
  author = {Artiukhov, Dmitrii and Brockbernd, Bob and Fedotova, Evgeniia and Koval, Nikita and Kylchik, Ivan and Moiseenko, Evgenii and Serebryakov, Lev and Zhelensky, Evgeniy and Zuev, Maksim},
  year = 2026,
  journal = {Journal of Object Technology},
  series = {{{DEBT}} 2025},
  volume = {25},
  number = {1},
  pages = {A9:1--5},
  publisher = {AITO},
  address = {Kaiserslautern, Germany},
  issn = {1660-1769},
  doi = {http://dx.doi.org/10.5381/jot.2026.25.1.a9},
  url = {https://www.jot.fm/issues/issue_2026_01/a9.pdf},
  langid = {english}
}

@techreport{BalzerExDAMSExtendableDebugging1969,
  type = {Memorandum},
  title = {{{ExDAMS}}: Extendable Debugging and Monitoring System},
  author = {Balzer, Robert M.},
  year = 1969,
  month = apr,
  number = {RM-5772-ARPA},
  pages = {42},
  address = {Santa Monica, CA, USA},
  institution = {The Rand Corporation},
  url = {https://www.rand.org/pubs/research_memoranda/RM5772.html},
  urldate = {2026-03-23},
  langid = {english}
}

@article{BianchiEtAlInlineVisualizationManipulation2024,
  title = {Inline Visualization and Manipulation of Real-Time Hardware Log for Supporting Debugging of Embedded Programs},
  author = {Bianchi, Andrea and Yap, Zhi Lin and Lertjaturaphat, Punn and Henley, Austin Z. and Moon, Kongpyung Justin and Kim, Yoonji},
  year = 2024,
  month = jun,
  journal = {Proceedings of the ACM on Human-Computer Interaction},
  volume = {8},
  number = {EICS},
  pages = {1--26},
  publisher = {ACM},
  address = {New York, NY, USA},
  issn = {2573-0142},
  doi = {10.1145/3660250},
  urldate = {2026-02-10},
  langid = {english}
}

@inproceedings{CleveZellerLocatingCausesProgram2005,
  title = {Locating Causes of Program Failures},
  booktitle = {Proceedings of the 27th {{International Conference}} on {{Software Engineering}}},
  author = {Cleve, Holger and Zeller, Andreas},
  year = 2005,
  series = {{{ICSE}} 2005},
  pages = {342--351},
  publisher = {ACM Press},
  address = {New York, NY, USA},
  doi = {10.1145/1062455.1062522},
  urldate = {2026-04-06},
  langid = {english}
}

@inproceedings{ConsensEtAlVisualizingQueryingDistributed1994,
  title = {Visualizing and Querying Distributed Event Traces with {{Hy}}+},
  booktitle = {Proceedings of the {{International Conference}} on {{Applications}} of {{Databases}} ({{ADB}} 1994)},
  author = {Consens, Mariano P. and Hasan, Masum Z. and Mendelzon, Alberto O.},
  editor = {Witold, Litwin and Risch, Tore},
  year = 1994,
  series = {Lecture {{Notes}} in {{Computer Science}}},
  volume = {819},
  publisher = {Springer-Verlag},
  address = {Berlin, Heidelberg},
  doi = {10.1007/3-540-58183-9_45},
  isbn = {978-3-540-48473-8}
}

@article{CornelissenEtAlControlledExperimentProgram2011,
  title = {A Controlled Experiment for Program Comprehension through Trace Visualization},
  author = {Cornelissen, Bas and Zaidman, Andy and {van Deursen}, Arie},
  year = 2011,
  month = may,
  journal = {IEEE Transactions on Software Engineering},
  volume = {37},
  number = {3},
  pages = {341--355},
  publisher = {IEEE},
  issn = {1939-3520},
  doi = {10.1109/TSE.2010.47},
  urldate = {2026-02-10}
}

@article{EisenstadtBrayshawTransparentPROLOGMachine1988,
  title = {The Transparent {{PROLOG}} Machine ({{TPM}}): An Execution Model and Graphical Debugger for Logic Programming},
  shorttitle = {The Transparent {{PROLOG}} Machine ({{TPM}})},
  author = {Eisenstadt, Marc and Brayshaw, Mike},
  year = 1988,
  month = dec,
  journal = {The Journal of Logic Programming},
  volume = {5},
  number = {4},
  pages = {277--342},
  publisher = {North-Holland},
  issn = {0743-1066},
  doi = {10.1016/0743-1066(88)90001-5},
  url = {https://www.sciencedirect.com/science/article/pii/0743106688900015},
  urldate = {2023-04-14},
  langid = {english}
}

@inproceedings{EngblomReviewReverseDebugging2012,
  title = {A Review of Reverse Debugging},
  booktitle = {Proceedings of the 2012 {{System}}, {{Software}}, {{SoC}} and {{Silicon Debug Conference}}},
  author = {Engblom, Jakob},
  year = 2012,
  month = sep,
  pages = {1--6},
  publisher = {IEEE},
  issn = {2114-3684},
  url = {https://ieeexplore.ieee.org/document/6338149},
  urldate = {2026-03-16},
  isbn = {978-2-9539987-5-7}
}

@inproceedings{EtheredgeCMeRunProgramLogic2004,
  title = {{{CMeRun}}: Program Logic Debugging Courseware for {{CS1}}/{{CS2}} Students},
  shorttitle = {{{CMeRun}}},
  booktitle = {Proceedings of the 35th {{ACM Technical Symposium}} on {{Computer Science Education}}},
  author = {Etheredge, Jim},
  year = 2004,
  month = mar,
  series = {{{SIGCSE}} 2004},
  pages = {22--25},
  publisher = {Association for Computing Machinery},
  address = {New York, NY, USA},
  doi = {10.1145/971300.971311},
  urldate = {2026-01-06},
  isbn = {978-1-58113-798-9}
}

@inproceedings{GoldsmithEtAlRelationalQueriesProgram2005,
  title = {Relational Queries over Program Traces},
  booktitle = {Proceedings of the 20th Annual {{ACM SIGPLAN}} Conference on {{Object-Oriented Programming}}, {{Systems}}, {{Languages}}, and {{Applications}}},
  author = {Goldsmith, Simon F. and O'Callahan, Robert and Aiken, Alex},
  year = 2005,
  month = oct,
  series = {{{OOPSLA}} 2005},
  pages = {385--402},
  publisher = {ACM},
  address = {New York, NY, USA},
  issn = {0362-1340},
  doi = {10.1145/1094811.1094841},
  urldate = {2025-12-18}
}

@inproceedings{GuoOnlinePythonTutor2013,
  title = {Online {{Python Tutor}}: Embeddable Web-Based Program Visualization for {{CS}} Education},
  shorttitle = {Online Python Tutor},
  booktitle = {Proceedings of the 44th {{ACM Technical Symposium}} on {{Computer Science Education}}},
  author = {Guo, Philip J.},
  year = 2013,
  month = mar,
  series = {{{SIGCSE}} 2013},
  pages = {579--584},
  publisher = {Association for Computing Machinery},
  address = {New York, NY, USA},
  doi = {10.1145/2445196.2445368},
  urldate = {2023-03-31},
  isbn = {978-1-4503-1868-6}
}

@inproceedings{GuoTenMillionUsers2021,
  title = {Ten Million Users and Ten Years Later: {{Python Tutor}}'s Design Guidelines for Building Scalable and Sustainable Research Software in Academia},
  shorttitle = {Ten Million Users and Ten Years Later},
  booktitle = {The 34th {{Annual ACM Symposium}} on {{User Interface Software}} and {{Technology}}},
  author = {Guo, Philip J.},
  year = 2021,
  month = oct,
  series = {{{UIST}} 2021},
  pages = {1235--1251},
  publisher = {Association for Computing Machinery},
  address = {New York, NY, USA},
  doi = {10.1145/3472749.3474819},
  urldate = {2023-03-31},
  isbn = {978-1-4503-8635-7}
}

@misc{HallBirdseye2025,
  title = {{{birdseye}}},
  author = {Hall, Alex},
  year = 2025,
  month = sep,
  url = {https://github.com/alexmojaki/birdseye},
  urldate = {2026-03-23}
}

@misc{HallSnoop2024,
  title = {{{snoop}}},
  author = {Hall, Alex},
  year = 2024,
  month = oct,
  url = {https://github.com/alexmojaki/snoop}
}

@article{HarroldEtAlEmpiricalInvestigationRelationship2000,
  title = {An Empirical Investigation of the Relationship between Spectra Differences and Regression Faults},
  author = {Harrold, Mary Jean and Rothermel, Gregg and Sayre, Kent and Wu, Rui and Yi, Liu},
  year = 2000,
  journal = {Software Testing, Verification and Reliability},
  volume = {10},
  number = {3},
  pages = {171--194},
  publisher = {John Wiley \& Sons, Ltd},
  issn = {1099-1689},
  doi = {10.1002/1099-1689(200009)10:3<171::AID-STVR209>3.0.CO;2-J},
  urldate = {2026-02-09},
  copyright = {Copyright \copyright{} 2000 John Wiley \& Sons, Ltd.},
  langid = {english}
}

@phdthesis{HeinsenEganAdvancedDebuggingProgram2015,
  title = {Advanced Debugging and Program Visualization for Novice {{C}} Programmers},
  author = {Heinsen Egan, Matthew},
  year = 2015,
  address = {Perth, Australia},
  url = {https://research-repository.uwa.edu.au/en/publications/advanced-debugging-and-program-visualization-for-novice-c-program/},
  urldate = {2026-01-06},
  langid = {english},
  school = {University of Western Australia}
}

@inproceedings{HoferEtAlDesignImplementationBackwardintime2006,
  title = {Design and Implementation of a Backward-in-Time Debugger},
  booktitle = {{{NODe}}/{{GSEM}} 2006},
  author = {Hofer, Christoph and Denker, Marcus and Ducasse, St{\'e}phane},
  year = 2006,
  series = {Lecture {{Notes}} in {{Informatics}}},
  volume = {P-88},
  pages = {17--32},
  publisher = {Gesellschaft f\"ur Informatik e.V.},
  url = {https://dl.gi.de/handle/20.500.12116/24100},
  urldate = {2026-03-27},
  isbn = {978-3-88579-182-9},
  langid = {english}
}

@inproceedings{HuangEtAlLEAPLightweightDeterministic2010,
  title = {{{LEAP}}: Lightweight Deterministic Multi-Processor Replay of Concurrent {{Java}} Programs},
  shorttitle = {{{LEAP}}},
  booktitle = {Proceedings of the 18th {{ACM SIGSOFT International Symposium}} on {{Foundations}} of {{Software Engineering}}},
  author = {Huang, Jeff and Liu, Peng and Zhang, Charles},
  year = 2010,
  month = nov,
  series = {{{SIGSOFT}}/{{FSE}} 2010},
  pages = {207--216},
  publisher = {ACM},
  address = {Santa Fe, NM, USA},
  doi = {10.1145/1882291.1882323},
  urldate = {2026-03-23},
  isbn = {978-1-60558-791-2},
  langid = {english}
}

@misc{INRIAToplevelSystemREPL2025,
  title = {The Toplevel System or {{REPL}} ({{OCaml}})},
  author = {{INRIA}},
  year = 2025,
  journal = {The OCaml Manual},
  url = {https://ocaml.org/manual/5.4/toplevel.html#s%3Atoplevel-directives},
  urldate = {2025-12-18},
  lastaccessed = {2025-12-18}
}

@inproceedings{JiangEtAlLogItSupportingProgramming2023,
  title = {Log-{{It}}: Supporting Programming with Interactive, Contextual, Structured, and Visual Logs},
  shorttitle = {Log-{{It}}},
  booktitle = {Proceedings of the 2023 {{CHI Conference}} on {{Human Factors}} in {{Computing Systems}}},
  author = {Jiang, Peiling and Sun, Fuling and Xia, Haijun},
  year = 2023,
  month = apr,
  series = {{{CHI}} 2023},
  pages = {1--16},
  publisher = {ACM},
  address = {New York, NY, USA},
  doi = {10.1145/3544548.3581403},
  urldate = {2026-02-10},
  isbn = {978-1-4503-9421-5},
  langid = {english}
}

@techreport{JolidonKappelerStudentreadyImplementationTracebased2025,
  type = {Semester Project},
  title = {A Student-Ready Implementation of Trace-Based Debugging for {{Java}}},
  author = {Jolidon, Bastien and Kappeler, Kelvin},
  year = 2025,
  month = jan,
  address = {Lausanne, Switzerland},
  institution = {EPFL}
}

@inproceedings{JonesHarroldEmpiricalEvaluationTarantula2005,
  title = {Empirical Evaluation of the Tarantula Automatic Fault-Localization Technique},
  booktitle = {Proceedings of the 20th {{IEEE}}/{{ACM International Conference}} on {{Automated Software Engineering}}},
  author = {Jones, James A. and Harrold, Mary Jean},
  year = 2005,
  month = nov,
  series = {{{ASE}} 2005},
  pages = {273--282},
  publisher = {ACM},
  address = {New York, NY, USA},
  doi = {10.1145/1101908.1101949},
  urldate = {2026-02-09},
  isbn = {978-1-58113-993-8},
  langid = {english}
}

@inproceedings{KangGuoOmnicodeNoviceOrientedLive2017,
  title = {Omnicode: A Novice-Oriented Live Programming Environment with Always-on Run-Time Value Visualizations},
  shorttitle = {Omnicode},
  booktitle = {Proceedings of the 30th {{Annual ACM Symposium}} on {{User Interface Software}} and {{Technology}}},
  author = {Kang, Hyeonsu and Guo, Philip J.},
  year = 2017,
  month = oct,
  series = {{{UIST}} 2017},
  pages = {737--745},
  publisher = {Association for Computing Machinery},
  address = {New York, NY, USA},
  doi = {10.1145/3126594.3126632},
  urldate = {2026-02-10},
  isbn = {978-1-4503-4981-9}
}

@mastersthesis{KappelerPrintWizardNextlevelTracebased2026,
  title = {{{PrintWizard}}: Next-Level Trace-Based Debugging},
  shorttitle = {{{PrintWizard}}},
  author = {Kappeler, Kelvin},
  year = 2026,
  month = jan,
  address = {Lausanne, Switzerland},
  url = {https://infoscience.epfl.ch/handle/20.500.14299/257909},
  urldate = {2026-04-07},
  langid = {english},
  school = {EPFL}
}

@inproceedings{KleynGingrichGraphTraceUnderstandingObjectoriented1988,
  title = {{{GraphTrace}}---Understanding Object-Oriented Systems Using Concurrently Animated Views},
  booktitle = {Proceedings of the 3rd {{Conference}} on {{Object-Oriented Programming Systems}}, {{Languages}}, and {{Applications}}},
  author = {Kleyn, Michael F. and Gingrich, Paul C.},
  year = 1988,
  series = {{{OOPSLA}} 1988},
  pages = {191--205},
  publisher = {Association for Computing Machinery},
  address = {New York, NY, USA},
  doi = {10.1145/62083.62101},
  urldate = {2026-02-10},
  isbn = {978-0-89791-284-6}
}

@inproceedings{KluyverEtAlJupyterNotebooksPublishing2016,
  title = {Jupyter {{Notebooks}} -- a Publishing Format for Reproducible Computational Workflows},
  booktitle = {Positioning and Power in Academic Publishing: Players, Agents and Agendas},
  author = {Kluyver, Thomas and {Ragan-Kelley}, Benjamin and P{\'e}rez, Fernando and Granger, Brian and Bussonnier, Matthias and Frederic, Jonathan and Kelley, Kyle and Hamrick, Jessica and Grout, Jason and Corlay, Sylvain and Ivanov, Paul and Avila, Dami{\'a}n and Abdalla, Safia and Willing, Carol and Jupyter Development Team},
  year = 2016,
  pages = {87--90},
  publisher = {IOS Press},
  address = {Amsterdam},
  doi = {10.3233/978-1-61499-649-1-87},
  url = {https://ebooks.iospress.nl/doi/10.3233/978-1-61499-649-1-87},
  urldate = {2023-03-31}
}

@inproceedings{KoEtAlAutoPrintJudgingEffectiveness2025,
  title = {{{AutoPrint}}: Judging the Effectiveness of an Automatic Print Statement Debugging Tool},
  shorttitle = {{{AutoPrint}}},
  booktitle = {2025 {{IEEE Symposium}} on {{Visual Languages}} and {{Human-Centric Computing}} ({{VL}}/{{HCC}})},
  author = {Ko, Minhyuk and Ahmed, Omer and Alebachew, Yoseph Berhanu and Brown, Chris},
  year = 2025,
  month = oct,
  pages = {379--384},
  publisher = {IEEE},
  issn = {1943-6106},
  doi = {10.1109/VL-HCC65237.2025.00049},
  urldate = {2026-01-14}
}

@inproceedings{KoMyersDebuggingReinventedAsking2008,
  title = {Debugging Reinvented: Asking and Answering Why and Why Not Questions about Program Behavior},
  shorttitle = {Debugging Reinvented},
  booktitle = {Proceedings of the 13th {{International Conference}} on {{Software Engineering}}},
  author = {Ko, Amy J. and Myers, Brad A.},
  year = 2008,
  series = {{{ICSE}} 2008},
  pages = {301--310},
  publisher = {ACM},
  address = {New York, NY, USA},
  doi = {10.1145/1368088.1368130},
  urldate = {2026-02-10},
  copyright = {https://www.acm.org/publications/policies/copyright\_policy\#Background},
  isbn = {978-1-60558-079-1},
  langid = {english}
}

@inproceedings{KoMyersDesigningWhylineDebugging2004,
  title = {Designing the {{Whyline}}: A Debugging Interface for Asking Questions about Program Behavior},
  shorttitle = {Designing the Whyline},
  booktitle = {Proceedings of the {{SIGCHI Conference}} on {{Human Factors}} in {{Computing Systems}}},
  author = {Ko, Amy J. and Myers, Brad A.},
  year = 2004,
  month = apr,
  series = {{{CHI}} 2004},
  pages = {151--158},
  publisher = {Association for Computing Machinery},
  address = {New York, NY, USA},
  doi = {10.1145/985692.985712},
  urldate = {2026-02-10},
  isbn = {978-1-58113-702-6}
}

@mastersthesis{LamperthRecordReplayDebugging2024,
  title = {Record and Replay Debugging at Scale: Towards Always-on {{Java}} Application Recording in the Cloud},
  shorttitle = {Record and {{Replay Debugging}} at {{Scale}}},
  author = {Lamp{\'e}rth, Jonathan},
  year = 2024,
  doi = {10.3929/ethz-b-000665219},
  url = {https://www.research-collection.ethz.ch/handle/20.500.11850/665219},
  urldate = {2025-06-04},
  copyright = {http://rightsstatements.org/page/InC-NC/1.0/},
  langid = {english},
  school = {ETH Zurich}
}

@article{LeDouxParkerSavingTracesAda1985a,
  title = {Saving Traces for {{Ada}} Debugging},
  author = {LeDoux, Carol H. and Parker, D. Stott},
  year = 1985,
  month = may,
  journal = {SIGAda Ada Letters},
  volume = {V},
  number = {2},
  pages = {97--108},
  publisher = {ACM},
  address = {New York, NY, USA},
  issn = {1094-3641},
  doi = {10.1145/324422.324385},
  urldate = {2026-03-24}
}

@inproceedings{LernerProjectionBoxesOnthefly2020,
  title = {Projection Boxes: On-the-Fly Reconfigurable Visualization for Live Programming},
  shorttitle = {Projection Boxes},
  booktitle = {Proceedings of the 2020 {{CHI Conference}} on {{Human Factors}} in {{Computing Systems}}},
  author = {Lerner, Sorin},
  year = 2020,
  month = apr,
  series = {{{CHI}} 2020},
  pages = {1--7},
  publisher = {Association for Computing Machinery},
  address = {New York, NY, USA},
  doi = {10.1145/3313831.3376494},
  urldate = {2026-02-10},
  isbn = {978-1-4503-6708-0}
}

@inproceedings{LewisDebuggingBackwardsTime2003,
  title = {Debugging {{Backwards}} in {{Time}}},
  booktitle = {Proceedings of the {{Fifth International Workshop}} on {{Automated}} and {{Algorithmic Debugging}}},
  author = {Lewis, Bil},
  year = 2003,
  month = sep,
  series = {{{AADEBUG}} 2003},
  eprint = {cs/0310016},
  pages = {225--235},
  publisher = {arXiv},
  address = {Ghent, Belgium},
  doi = {10.48550/arXiv.cs/0310016},
  url = {http://arxiv.org/abs/cs/0310016},
  urldate = {2026-03-16},
  archiveprefix = {arXiv}
}

@inproceedings{LieberEtAlAddressingMisconceptionsCode2014,
  title = {Addressing Misconceptions about Code with Always-on Programming Visualizations},
  booktitle = {Proceedings of the {{SIGCHI Conference}} on {{Human Factors}} in {{Computing Systems}}},
  author = {Lieber, Tom and Brandt, Joel R. and Miller, Rob C.},
  year = 2014,
  month = apr,
  series = {{{CHI}} 2014},
  pages = {2481--2490},
  publisher = {ACM},
  address = {New York, NY, USA},
  doi = {10.1145/2556288.2557409},
  urldate = {2026-02-10},
  isbn = {978-1-4503-2473-1},
  langid = {english}
}

@inproceedings{LiebermanStepsBetterDebugging1984,
  title = {Steps toward Better Debugging Tools for {{LISP}}},
  booktitle = {Proceedings of the 1984 {{ACM Symposium}} on {{LISP}} and {{Functional Programming}} ({{LFP}} 1984)},
  author = {Lieberman, Henry},
  year = 1984,
  month = aug,
  pages = {247--255},
  publisher = {ACM},
  address = {Austin, TX, USA},
  doi = {10.1145/800055.802041},
  urldate = {2026-03-24},
  copyright = {https://www.acm.org/publications/policies/copyright\_policy\#Background},
  isbn = {978-0-89791-142-9},
  langid = {english}
}

@misc{LieberTheseus2015,
  title = {Theseus},
  author = {Lieber, Tom},
  year = 2015,
  url = {https://github.com/adobe-research/theseus},
  urldate = {2026-02-10},
  copyright = {MIT}
}

@misc{LispWorksLtd.MacroTRACEUNTRACE2025,
  title = {Macro {{TRACE}}, {{UNTRACE}}},
  author = {{LispWorks Ltd.}},
  year = 2025,
  journal = {Common Lisp HyperSpec},
  url = {http://clhs.lisp.se/Body/m_tracec.htm},
  urldate = {2025-12-18},
  lastaccessed = {2025-12-18}
}

@inproceedings{MartinEtAlFindingApplicationErrors2005,
  title = {Finding Application Errors and Security Flaws Using {{PQL}}: A Program Query Language},
  booktitle = {Proceedings of the 20th Annual {{ACM SIGPLAN}} Conference on {{Object-Oriented Programming}}, {{Systems}}, {{Languages}}, and {{Applications}}},
  author = {Martin, Michael and Livshits, Benjamin and Lam, Monica S.},
  year = 2005,
  month = oct,
  series = {{{OOPSLA}} 2005},
  pages = {365--383},
  publisher = {ACM},
  address = {San Diego, CA, USA},
  doi = {10.1145/1094811.1094840},
  urldate = {2026-03-24},
  isbn = {978-1-59593-031-6},
  langid = {english}
}

@inproceedings{MatsumuraEtAlRepeatedlyexecutedmethodViewerEfficient2014,
  title = {Repeatedly-Executed-Method Viewer for Efficient Visualization of Execution Paths and States in {{Java}}},
  booktitle = {Proceedings of the 22nd {{International Conference}} on {{Program Comprehension}}},
  author = {Matsumura, Toshinori and Ishio, Takashi and Kashima, Yu and Inoue, Katsuro},
  year = 2014,
  month = jun,
  series = {{{ICPC}} 2014},
  pages = {253--257},
  publisher = {Association for Computing Machinery},
  address = {New York, NY, USA},
  doi = {10.1145/2597008.2597803},
  urldate = {2025-06-04},
  isbn = {978-1-4503-2879-1}
}

@misc{MicrosoftCorp.IntelliTrace2025,
  title = {{{IntelliTrace}}},
  author = {{Microsoft Corp.}},
  year = 2025,
  month = oct,
  url = {https://learn.microsoft.com/en-us/visualstudio/debugger/intellitrace?view=visualstudio},
  urldate = {2026-02-10},
  howpublished = {Microsoft Corporation}
}

@misc{MonettaFlowStorm2025,
  title = {{{FlowStorm}}},
  author = {Monetta, Juan},
  year = 2025,
  month = sep,
  url = {https://www.flow-storm.org/},
  urldate = {2026-03-16}
}

@inproceedings{MurphyEtAlBackstopToolDebugging2008,
  title = {Backstop: A Tool for Debugging Runtime Errors},
  shorttitle = {Backstop},
  booktitle = {Proceedings of the 39th {{ACM Technical Symposium}} on {{Computer Science Education}}},
  author = {Murphy, Christian and Kim, Eunhee and Kaiser, Gail and Cannon, Adam},
  year = 2008,
  month = mar,
  series = {{{SIGCSE}} '08},
  pages = {5},
  publisher = {Association for Computing Machinery},
  address = {New York, NY, USA},
  doi = {10.1145/1352135.1352193},
  isbn = {978-1-59593-799-5},
  langid = {english}
}

@techreport{OCallahanEtAlEngineeringRecordReplay2017,
  title = {Engineering Record and Replay for Deployability: Extended Technical Report},
  shorttitle = {Engineering {{Record And Replay For Deployability}}},
  author = {O'Callahan, Robert and Jones, Chris and Froyd, Nathan and Huey, Kyle and Noll, Albert and Partush, Nimrod},
  year = 2017,
  month = may,
  eprint = {1705.05937},
  primaryclass = {cs},
  doi = {10.48550/arXiv.1705.05937},
  url = {http://arxiv.org/abs/1705.05937},
  urldate = {2025-06-04},
  archiveprefix = {arXiv}
}

@misc{PernoscoPernosco2026,
  title = {Pernosco},
  author = {{Pernosco}},
  year = 2026,
  url = {https://pernos.co/},
  urldate = {2026-03-16},
  howpublished = {Pernosco}
}

@inproceedings{Pit-ClaudelUntanglingMechanizedProofs2020,
  title = {Untangling Mechanized Proofs},
  booktitle = {Proceedings of the 13th {{ACM SIGPLAN International Conference}} on {{Software Language Engineering}}},
  author = {{Pit-Claudel}, Cl{\'e}ment},
  year = 2020,
  month = nov,
  volume = {6167},
  pages = {155--174},
  publisher = {ACM},
  address = {Virtual USA},
  doi = {10.1145/3426425.3426940},
  urldate = {2023-07-13},
  isbn = {978-1-4503-8176-5},
  langid = {english}
}

@misc{PosmaKrouseLessons2014,
  title = {{$\lambda$} {{Lessons}}},
  author = {Posma, Jan Paul and Krouse, Steve},
  year = 2014,
  month = aug,
  url = {https://stevekrouse.com/hs.js/},
  urldate = {2026-03-16}
}

@article{PothierTanterBackFutureOmniscient2009,
  title = {Back to the Future: Omniscient Debugging},
  shorttitle = {Back to the {{Future}}},
  author = {Pothier, Guillaume and Tanter, {\'E}ric},
  year = 2009,
  month = nov,
  journal = {IEEE Software},
  volume = {26},
  number = {6},
  pages = {78--85},
  issn = {1937-4194},
  doi = {10.1109/MS.2009.169},
  urldate = {2025-06-04}
}

@misc{RachumEtAlPySnooperNeverUse2019,
  title = {{{PySnooper}}: Never Use {{print}} for Debugging Again},
  author = {Rachum, Ram and Hall, Alex and Yanokura, Iori and {PySnooper contributors}},
  year = 2019,
  month = jun,
  doi = {10.5281/zenodo.10462459},
  url = {https://github.com/cool-RR/PySnooper},
  howpublished = {PyCon Israel}
}

@misc{RecordReplayInc.Replay2026,
  title = {Replay},
  author = {{Record Replay Inc.}},
  year = 2026,
  url = {https://www.replay.io/},
  urldate = {2026-04-02}
}

@inproceedings{RepsEtAlUseProgramProfiling1997,
  title = {The Use of Program Profiling for Software Maintenance with Applications to the Year 2000 Problem},
  booktitle = {Proceedings of the 6th {{European Software Engineering Conference}}, Held Jointly with the 5th {{ACM SIGSOFT International Symposium}} on {{Foundations}} of {{Software Engineering}}},
  author = {Reps, Thomas and Ball, Thomas and Das, Manuvir and Larus, James},
  year = 1997,
  month = nov,
  series = {{{ESEC}} 1997/{{FSE-5}}},
  pages = {432--449},
  publisher = {Springer-Verlag},
  address = {Berlin, Heidelberg},
  doi = {10.1145/267895.267925},
  urldate = {2026-02-09},
  isbn = {978-3-540-63531-4}
}

@article{SakuraiEtAlTraceglassesTracebasedDebugger2010,
  title = {{Traceglasses: a trace-based debugger for realizing efficient navigation}},
  author = {Sakurai, Kouhei and Masuhara, Hidehiko and Komiya, Seiichi},
  year = 2010,
  month = jun,
  journal = {IPSJ Transactions on Programming},
  volume = {3},
  number = {3},
  pages = {1--17},
  publisher = {Information Processing Society of Japan},
  url = {https://prg.is.titech.ac.jp/papers/pdf/ipsj-trans-pro-2010.pdf},
  urldate = {2026-02-10},
  langid = {jp}
}

@article{SchwartzEtAlJmvxFastMultithreaded2024,
  title = {Jmvx: Fast Multi-Threaded Multi-Version Execution and Record-Replay for Managed Languages},
  shorttitle = {Jmvx},
  author = {Schwartz, David and Kowshik, Ankith and Pina, Lu{\'i}s},
  year = 2024,
  month = oct,
  journal = {Proceedings of the ACM on Programming Languages},
  series = {{{OOPSLA}} 2024},
  volume = {8},
  number = {OOPSLA2},
  pages = {1641--1669},
  issn = {2475-1421},
  doi = {10.1145/3689769},
  urldate = {2026-03-23},
  langid = {english}
}

@mastersthesis{SerandourImmediateTracingSmoother2024,
  title = {Immediate Tracing for Smoother Debugging and Code Exploration},
  author = {Serandour, Erwan},
  year = 2024,
  month = jun,
  address = {Lausanne, Switzerland},
  school = {EPFL}
}

@inproceedings{SotoudehLiterateTracing2025,
  title = {Literate Tracing},
  booktitle = {Proceedings of the 2025 {{ACM SIGPLAN International Symposium}} on {{New Ideas}}, {{New Paradigms}}, and {{Reflections}} on {{Programming}} and {{Software}}},
  author = {Sotoudeh, Matthew},
  year = 2025,
  month = oct,
  pages = {143--160},
  publisher = {Association for Computing Machinery},
  address = {New York, NY, USA},
  doi = {10.1145/3759429.3762626},
  urldate = {2025-12-18},
  isbn = {979-8-4007-2151-9},
  langid = {english}
}

@inproceedings{SuzukiEtAlTraceDiffDebuggingUnexpected2017,
  title = {{{TraceDiff}}: {{Debugging}} Unexpected Code Behavior Using Trace Divergences},
  shorttitle = {{{TraceDiff}}},
  booktitle = {2017 {{IEEE Symposium}} on {{Visual Languages}} and {{Human-Centric Computing}} ({{VL}}/{{HCC}})},
  author = {Suzuki, Ryo and Soares, Gustavo and Head, Andrew and Glassman, Elena and Reis, Ruan and Mongiovi, Melina and D'Antoni, Loris and Hartmann, Bj{\"o}rn},
  year = 2017,
  month = oct,
  pages = {107--115},
  publisher = {IEEE},
  issn = {1943-6106},
  doi = {10.1109/VLHCC.2017.8103457},
  urldate = {2026-02-10}
}

@misc{SWIPrologDebuggingTracingPrograms2026,
  title = {Debugging and Tracing Programs},
  author = {{SWI Prolog}},
  year = 2026,
  journal = {SWI Prolog reference manual},
  url = {https://www.swi-prolog.org/pldoc/man?section=debugger},
  urldate = {2026-03-27},
  lastaccessed = {2026-03-27}
}

@inproceedings{TankinkEtAlProviolaToolProof2010,
  title = {Proviola: A Tool for Proof Re-Animation},
  shorttitle = {Proviola},
  booktitle = {Intelligent {{Computer Mathematics}} ({{CICM}} 2010)},
  author = {Tankink, Carst and Geuvers, Herman and McKinna, James and Wiedijk, Freek},
  editor = {Autexier, Serge and Calmet, Jacques and Delahaye, David and Ion, Patrick D. F. and Rideau, Laurence and Rioboo, Renaud and Sexton, Alan P.},
  year = 2010,
  series = {Lecture {{Notes}} in {{Computer Science}}},
  number = {6167},
  pages = {440--454},
  publisher = {Springer},
  address = {Berlin, Heidelberg},
  doi = {10.1007/978-3-642-14128-7_37},
  isbn = {978-3-642-14128-7},
  langid = {english}
}

@inproceedings{TeradaETVProgramTrace2005,
  title = {{{ETV}}: A Program Trace Player for Students},
  shorttitle = {{{ETV}}},
  booktitle = {Proceedings of the 10th {{Annual SIGCSE Conference}} on {{Innovation}} and {{Technology}} in {{Computer Science Education}}},
  author = {Terada, Minoru},
  year = 2005,
  month = jun,
  series = {{{ITiCSE}} 2005},
  pages = {118--122},
  publisher = {Association for Computing Machinery},
  address = {New York, NY, USA},
  doi = {10.1145/1067445.1067480},
  urldate = {2026-01-06},
  isbn = {978-1-59593-024-8},
  langid = {english}
}

@misc{UndoLtd.UndoJava2026,
  title = {Undo for {{Java}}},
  author = {{Undo Ltd.}},
  year = 2026,
  url = {https://undo.io/products/java/},
  urldate = {2026-03-23}
}

@misc{VasconcelosHaskelite2025,
  title = {Haskelite},
  author = {Vasconcelos, Pedro Baltazar},
  year = 2025,
  month = sep,
  url = {https://github.com/pbv/haskelite}
}

@inproceedings{Velazquez-IturbideEtAlSRecAnimationSystem2008,
  title = {{{SRec}}: An Animation System of Recursion for Algorithm Courses},
  shorttitle = {{{SRec}}},
  booktitle = {Proceedings of the 13th {{Annual SIGCSE Conference}} on {{Innovation}} and {{Technology}} in {{Computer Science Education}}},
  author = {{Vel{\'a}zquez-Iturbide}, J. {\'A}ngel and {P{\'e}rez-Carrasco}, Antonio and {Urquiza-Fuentes}, Jaime},
  year = 2008,
  month = jun,
  series = {{{ITiCSE}} '08},
  pages = {225--229},
  publisher = {Association for Computing Machinery},
  address = {New York, NY, USA},
  doi = {10.1145/1384271.1384332},
  urldate = {2026-01-06},
  isbn = {978-1-60558-078-4}
}

@misc{VictorLearnableProgramming2012,
  title = {Learnable Programming},
  author = {Victor, Bret},
  year = 2012,
  month = sep,
  journal = {Bret Victor, human being},
  url = {https://worrydream.com/LearnableProgramming/},
  urldate = {2026-04-07},
  lastaccessed = {2026-04-07}
}

@inproceedings{WangLaTozaHowOmniscientDebuggers2025,
  title = {How Omniscient Debuggers Impact Debugging Behavior},
  booktitle = {2025 {{IEEE Symposium}} on {{Visual Languages}} and {{Human-Centric Computing}} ({{VL}}/{{HCC}})},
  author = {Wang, Ruochen and LaToza, Thomas D.},
  year = 2025,
  month = oct,
  pages = {57--67},
  publisher = {IEEE},
  issn = {1943-6106},
  doi = {10.1109/VL-HCC65237.2025.00016},
  urldate = {2026-02-10}
}

@inproceedings{WeningerEtAlJavaWizTracebasedGraphical2025,
  title = {{{JavaWiz}}: A Trace-Based Graphical Debugger for Software Development Education},
  shorttitle = {{{JavaWiz}}},
  booktitle = {2025 {{IEEE}}/{{ACM}} 33rd {{International Conference}} on {{Program Comprehension}} ({{ICPC}})},
  author = {Weninger, Markus and Gr{\"u}nbacher, Simon and Pr{\"a}hofer, Herbert},
  year = 2025,
  month = apr,
  pages = {1--12},
  publisher = {IEEE},
  address = {Ottawa, ON, Canada},
  issn = {2643-7171},
  doi = {10.1109/ICPC66645.2025.00023},
  urldate = {2025-08-05}
}

@article{WongEtAlDStarMethodEffective2014,
  title = {The {{DStar}} Method for Effective Software Fault Localization},
  author = {Wong, W. Eric and Debroy, Vidroha and Gao, Ruizhi and Li, Yihao},
  year = 2014,
  month = mar,
  journal = {IEEE Transactions on Reliability},
  volume = {63},
  number = {1},
  pages = {290--308},
  publisher = {IEEE},
  issn = {1558-1721},
  doi = {10.1109/TR.2013.2285319},
  urldate = {2026-02-09}
}

@book{ZellerWhyProgramsFail2009,
  title = {Why Programs Fail: A Guide to Systematic Debugging},
  author = {Zeller, Andreas},
  year = 2009,
  month = jun,
  edition = {2nd},
  publisher = {Morgan Kaufmann Publishers Inc.},
  address = {San Francisco, CA, USA},
  isbn = {978-0-12-374515-6}
}

@inproceedings{ZhaoEtAlUnderstandingTriggerActionPrograms2021,
  title = {Understanding Trigger-Action Programs through Novel Visualizations of Program Differences},
  booktitle = {Proceedings of the 2021 {{CHI Conference}} on {{Human Factors}} in {{Computing Systems}}},
  author = {Zhao, Valerie and Zhang, Lefan and Wang, Bo and Littman, Michael L. and Lu, Shan and Ur, Blase},
  year = 2021,
  month = may,
  series = {{{CHI}} 2021},
  pages = {1--17},
  publisher = {Association for Computing Machinery},
  address = {New York, NY, USA},
  doi = {10.1145/3411764.3445567},
  urldate = {2026-02-10},
  isbn = {978-1-4503-8096-6}
}

@article{ZhongEtAlLogUpdaterAutomatedDetection2025,
  title = {{{LogUpdater}}: Automated Detection and Repair of Specific Defects in Logging Statements},
  shorttitle = {{{LogUpdater}}},
  author = {Zhong, Renyi and Li, Yichen and Kuang, Jinxi and Gu, Wenwei and Huo, Yintong and Lyu, Michael R.},
  year = 2025,
  month = dec,
  journal = {ACM Transactions on Software Engineering and Methodology},
  volume = {35},
  number = {1},
  pages = {16:1--16:31},
  publisher = {ACM},
  address = {New York, NY, USA},
  issn = {1049-331X},
  doi = {10.1145/3731754},
  urldate = {2026-04-06}
}

\end{document}